\documentstyle[12pt,epsf]{article}
\begin{document}
\title{
\rightline{\small  KANAZAWA--98/07}
\vspace{-10pt}
\rightline{\small  HUB--EP--98/64}
\vspace{-10pt}
\rightline{\small ITEP 49/98}
\vspace{1cm}
Background Configurations, Confinement
and Deconfinement on a Lattice 
with BPS Monopole Boundary Conditions
}
\author{
E.-M. Ilgenfritz$^1$,
S.V. Molodtsov$^2$,
M. M\"uller-Preussker$^3$,  \\
and A.I. Veselov$^2$  \\
{\small{\it $^1$ Institute for Theoretical Physics,
             University of Kanazawa, Japan}} \\
{\small{\it $^2$ Institute of Theoretical and Experimental
             Physics, Moscow, Russia}} \\
{\small{\it $^3$ Institut f\"ur Physik,
             Humboldt-Universit\"at zu Berlin, Germany}}
}
\date{September 25, 1998}
\maketitle

\begin{abstract}
Finite temperature $SU(2)$ lattice gauge theory is investigated in a
$3D$ cubic box with fixed boundary conditions provided by a
discretized,
static BPS monopole solution with varying core scale ${\mu}$.
Using heating and cooling techniques we establish that for discrete
${\mu}$-values stable classical
solutions either of self-dual or of pure magnetic
type exist inside the box. Having switched on quantum fluctuations
we compute the Polyakov line and
other local operators. For different ${\mu}$ and at varying temperatures
near the deconfinement transition we study the influence of the boundary
condition on the vacuum inside the box. In contrast to the
pure magnetic background field case, for the self-dual one
we observe confinement even for temperatures quite far
above the critical one.
\end{abstract}

\newpage
\section{Introduction}
Quark confinement is one of the most intriguing problems in QCD.
There are several appealing analogies to SQCD and supergravity which
are helpful to guide our intuition. Nevertheless, a final
explanation of the mechanism in terms of structures of
(non-supersymmetric) QCD or pure Yang-Mills theory itself is still missing.
More precisely, it is confinement at weak coupling, when the lattice spacing
is well below the presumably relevant structures accomplishing this
mechanism, that we need to explain. Therefore a closer investigation of
various semi-classical configurations in the context of lattice simulations
in different phases seems to be necessary.

The first example of an extended excitation which is looked for,
nowadays massively, in Monte Carlo generated gauge field configurations
is the instanton \cite{inst}. It is not completely clear at present
to what extent their size (actually their distribution of sizes) is
influenced by the detection algorithm itself \cite{inst_on_lattice,negele98},
but there is no doubt
that this scale can be and perhaps has been already decoupled from the
lattice spacing.
\footnote{Perfect actions, however, are trying to go the opposite way
\cite{perf_act}. At the cost of a complicated but still practically
feasible action one wants to be able to keep the discretization scale near
to the dynamically important scale(s) whereas good continuum features can be
obtained within bigger volumes in affordable computer time.}

In a recent publication lattice measurements of the field strength
correlator at zero temperature have been confronted with semi-analytic
results of the semi-classically motivated instanton liquid
approximation \cite{instcorr}.
The correlation length and strength of the correlation at intermediate
distances can be explained within the standard parameters characterizing
the instanton liquid. However, the change of the field strength
correlators at the deconfining transition cannot be explained in an model
based exclusively on self-dual or anti-self-dual configurations \cite{calcorr}.

According to current folklore, instantons are not related to confinement
at all. If they are related, we have still to identify the
interactions and correlations in the instanton liquid which
would make the corresponding sample fields confining. Traditionally, in
continuum models, instantons are the relevant configurations to explain
spontaneous chiral symmetry breaking and are needed to solve the
$~U_A(1)~$ problem \cite{inst}.
Without the inclusion of (short range) instanton correlations one is
not able to unterstand the temperature dependence of the chiral
condensates and the restoration of chiral symmetry at $T_{ch}$
(for a schematic model, see Ref. \cite{ilgshur}).
An indirect role for confinement is attributed to instantons
because of local correlations between monopoles and instantons
\cite{inst_mon}.

Monopoles are the agents of another model that has been first
formulated in the continuum in order to explain confinement:
the dual superconductor scenario invented by t'Hooft and Mandelstam
\cite{mon1}. It assumes condensation of Abelian monopoles and views
confinement as a dual Meissner effect. That this scenario really
takes place has been demonstrated for Abelian compact and
non-Abelian lattice gauge theories by numerical simulations \cite{mon2,mon3}.
In the latter case Abelian monopoles appear as singularities
of the gauge fixing procedure and are identified as point-like
conserved currents (DeGrand-Toussaint monopoles \cite{degrand})
in various gauges after Abelian projection. These monopoles
have a quantized magnetic charge but no natural size which would be
analogous to the instanton size and could point to some confinement
scale and corresponding QCD coupling. In the very first attempt it was
tried to prove that their {\it density} possesses a continuum limit
\cite{bornyakov}. But the density is strongly affected by short-range
lattice artefacts. There have been methods developed to remove these
artefacts (monopole blocking to obtain an infrared effective action for
thick monopoles \cite{suzuki}, renormalization group motivated smoothing
of configurations which reduces point-like Abelian monopole currents
\cite{smoothing}).

There is one model for an extended, particle-like monopole which could
be confronted with first-principle lattice Monte Carlo configurations.
This is the t'Hooft-Polyakov (HP) \cite{thp} or -- as the limiting case --
the Bogomol'nyi-Prasad-Sommerfield (BPS) monopole  \cite{bps}.
It is taken from the Georgi-Glashow model and re-interpreted as a
static configuration in pure gauge theory identifying
the adjoint Higgs field with the $4^{\mbox{th}}$ component
of the vector potential.
As a solution of the classical non-Abelian field equations the BPS monopole
satisfies the self-duality condition and carries electric charge.
This is the reason why in pure gauge theory we want to call the BPS
monopole solution also {\it dyon}.
\footnote{One should not mistake this dyon solution with
the Julia-Zee dyon in the Georgi-Glashow model \cite{juliazee}.}

The present paper represents an attempt to examine isolated
configurations of this kind under the influence of quantum fluctuations.
A first investigation in this spirit was reported by Smit and van der Sijs
\cite{smitvds1}.  A particular feature in our present work
(similar as in \cite{smitvds1})
is the possibility to select the semi-classical background field by
applying specific boundary conditions fixing the tangential vector
potential at the surface of a spatial box. The main emphasis in the
work of our precedessors was to select the coupling (or length scale)
at which these monopoles would condense at zero physical temperature
(and to determine the mass of the quantum monopole) in the spirit of
Ref. \cite{smitvds2}. Here, in contrast, we want first to elaborate
on the question which solutions of the lattice equations of motion
are compatible with the boundary  conditions. Only if they are
invariably recovered from heating and cooling cycles
we can speak of unique background configurations. We will
discover that, although the boundary conditions are
originally taken from a self-dual BPS configuration, depending on
discrete values of the size parameter $\mu^{-1}$, these
solutions can have either pure magnetic monopole or dyonic (self-dual)
character.
For simplicity, let us call them in the first case {\it HP monopole}
(because of the suppressed electric components),
in the second one {\it BPS monopole} or simply {\it dyon}.
Our main interest is to investigate for the finite temperature case
how usual indicators of the deconfinement phase transition
(average Polyakov line, distribution of Polyakov lines) are
modified by the presence of the different kinds of background fields.
For this purpose we have to consider these observables locally,
sufficiently far from the `cold walls' of the box.

Our main result will be that the self-dual BPS-like (dyon)
background fields -- in contrast to
the pure magnetic HP monopole ones -- strongly support confinement.
We shall show that a dyon invironment keeps the
Yang-Mills theory in the confinement state even for temperatures
$~T > T_c~$.

From the investigation of the field strength correlator \cite{instcorr}
it is known that self-dual semi-classical configurations seem to be essential
in the confinement phase {\it but only there}. Investigations of
Wilson loops in the classical dyon background have shown that
dyons give rise, at short distances, to a weak but confining quark-antiquark
force of harmonic oscillator type ('super-confinement') \cite{superconf}.
\footnote{Similar observations have been made for the short range potential
in the field of an instanton \cite{instpot} and for more generic self-dual
fields \cite{efimov}.}
Notice that this is not the confining force at long distances described
by the string tension or the magnetic confinement \cite{simonov} for
space-like Wilson loops at high temperature.
With respect to the latter, a dense gas of magnetic monopoles at high
temperatures could explain it.

The paper is organized as follows. In chapter 2 we show how to
discretize a continuum BPS monopole (dyon) solution
and how the classical lattice configurations compatible with dyonic
boundary conditions can be classified interpolating between lattice
HP monopoles and dyons. In chapter 3 we report on Monte Carlo
simulations performed with fixed open spatial boundary conditions
corresponding to HP monopole or dyon background fields
at finite temperature. We measure the profile of the
action density and Polyakov line inside the box
and establish the existence or absence of confinement.
Chapter 4 contains the conclusions.

\section{Classical Solutions with Dyon Boundary \\ Conditions}
First let us discuss classical field configurations in a finite volume
with fixed spatial boundary conditions.
We consider a hypercubic lattice of size $N_s^3 \times N_t$ with
periodic boundary conditions in the imaginary time direction (finite
temperature $T = 1/(a \cdot N_t)$)
and open boundary conditions in the three-space. The latter
will be specified such that links
being normal to the boundary and pointing outwards are
inactive, i. e. cannot contribute to the
action. Link variables tangential to the boundaries will be fixed to
classical values given by a BPS dyon configuration as defined
in the following.

We discretize a single BPS dyon field being a static solution
of the Euclidean Yang-Mills equations of motion. Its center
be fixed at $x_1=x_2=x_3=(N_s+1)/2$. For
the lattice spacing $~a=1~$ is assumed.
On each time slice $~t \equiv x_4~$ all active link variables
$~U_{\nu}(\vec{x},t) ~\in SU(2)$, $\nu=1,..,4~$,
are represented as follows:
\begin{eqnarray}\label{dyon1} \nonumber
U_k(\vec{x},t) & = & \exp \left( \frac{-i}{2}
                        \sigma_j \varepsilon_{jkl} x_l
               \int_0^1 ds \frac{1 - K(\mu r(s))}{r^2(s)}
                         \right) \\
U_4(\vec{x},t) & = & \exp \left( \frac{-i}{2}
                         \sigma_j x_j \frac{H(\mu r)}{r}
                         \right),  \qquad \qquad j,k,l = 1,2,3  \,\, ,
\end{eqnarray}
with
$~r(s) = \sqrt{\left(x_k + s \right)^2 + \sum_{j \ne k}x_j^2}\,, \quad
r(0) = r~.$
For the BPS dyon solution we have explicitely
\begin{eqnarray}\label{dyon2}  \nonumber
K(\mu r) & = & \frac{\mu r}{sinh(\mu r)} \\
H(\mu r) & = & \mu r \frac{cosh(\mu r)}{sinh(\mu r)} - 1\,.
\end{eqnarray}
$\mu$ denotes the inverse core size of the dyon to be fixed lateron.
The replacement
$~\mu \rightarrow -\mu~$
transforms the dyon into an anti-dyon.
Note that in the limit $~r \rightarrow \infty~$
\begin{equation}
\frac{1}{r} H(\mu r) \sim \mu - \frac{1}{r} - O(\exp(-\mu r)).
\end{equation}
For comparison, the asymptotic behaviour of an HP pure monopole solution
(which classically would be related to a sufficiently strong
non-vanishing Higgs potential and in the quantum case of pure $SU(2)$
gauge theory could be due to a dynamical screening of the electric
field components) is determined by
\begin{equation} \label{monopole}
\frac{1}{r} H(\mu r) \sim \mu - O(\exp(-\mu r))\,.
\end{equation}
It is immediately clear that at semi-asymptotic distances
one cannot distinguish between a classical BPS dyon and a HP monopole
if one replaces
\begin{equation} \label{dyon_vs_mon}
 \mu_{dyon} - \frac{1}{r} \rightarrow \mu_{monopole} \,.
\end{equation}
As has been shown by Smit and van der Sijs \cite{smitvds1} the dyon or
monopole fields at the boundary have some further interesting properties.
Additionally to the Abelian gauge symmetry with respect to transformations
\begin{equation} \label{abgauge}
\Omega(\vec{x}, t) = \exp \left( \frac{i \omega}{2}
                     \sigma_j x_j  \right)
\end{equation}
($\omega = \mbox{const.}$)
the boundary fields periodic in $~t~$ exhibit a combined symmetry with
respect to the gauge transformation
\begin{equation} \label{combgauge}
\Omega(\vec{x}, t) = \exp \left( i \pi n \frac{t}{N_t}
                     \sigma_j x_j  \right)
\end{equation}
and the shift
\begin{equation} \label{mu_period}
 \mu \rightarrow \mu + \frac{2 \pi n}{N_t} \,, \quad n=1,2,...
\end{equation}

Both the shifts (\ref{dyon_vs_mon}) and (\ref{mu_period})
can be combined and let us expect a periodic behaviour in $\mu$ which
allows to create boundary conditions compatible with a BPS dyon
as well as with a HP monopole.
As far as on the hypercubic lattice we have no cylindric symmetry
at the boundary we have to replace $~r(\vec{x})~$ by an effective
$~R_{eff}~$, for which we adopt the value
$~R_{eff} \simeq 1.13 \cdot \frac12 (N_s-1)~$ as given in \cite{smitvds1}.
In what follows we shall parametrize the boundary values by
\begin{equation} \label{mu_prime}
 \mu' = (\mu - \frac{1}{R_{eff}}) \cdot \frac{N_t}{2 \pi}
\end{equation}
instead of $\mu$.

In general, one cannot expect the lattice discretized dyon to be a solution
of the {\it lattice equations of motion}. In order to find such a solution
corresponding to the dyon boundary conditions
we keep the tangential boundary links  fixed
at the classical values given by (\ref{dyon1}, \ref{dyon2}).
For varying $~\mu~$ or $~\mu'~$, respectively,
the other links inside the box (for all time slices) are exposed to
repeated cycles of Monte Carlo heating followed by cooling.
These links possess the form (\ref{dyon1}, \ref{dyon2}) only as start values.
We wanted to find out, what lattice fields, respecting the boundary
conditions augmented by periodicity in time, have minimal action.
It turned out that there exist {\it several
local} minima of action for $~\mu'~$ sufficiently big.
The configurations with lowest action show the interesting
$~\mu'$-dependence anticipated above.

In order to characterize the lattice field configurations obtained
by cooling after heating the discretized BPS dyon 
we will use the full plaquette action
\begin{equation}\label{action}
S = \beta \sum_{x,\mu < \nu} \left( 1 - \frac12 Tr U_{\mu,\nu}(x) \right)
   = E^2 + B^2\,, \qquad \beta \equiv 4/g^2\,,
\end{equation}
with its magnetic part $B^2$ coming from the sum of space-space 
plaquette contributions
and its electric part $E^2$  obtained as the sum of time-like plaquettes.
Fig. 1 shows the full action per timeslice of the
classical continuum BPS dyon, lattice discretized according
to (\ref{dyon1}, \ref{dyon2}), in dependence
on the parameter $\mu'$ (solid line). The action values are given in
units of $4 \pi / (a~g^2)$.
The lattice size is $12^3 \times 4$.
Obviously, for $~\mu' \simeq 2 ~$ the dyon core parameter becomes
sufficiently large such that this dyon completely fits into the
lattice box.

The data points in Fig. 1 show the global or local minima,
respectively, found for the full action (asterisks) after cooling down
the Monte Carlo heated configurations. 
Only in the range $~0 \le \mu' \le 0.5~$ (i.e. for a sufficiently
smooth discretized original dyon field
(\ref{dyon1}, \ref{dyon2})) the action is practically reproduced.
For larger $~\mu'$-values (less than 1.5) the solutions of the
lattice equations of motion
deviate from the original discretized dyon but are still unique
in the sense that they are restored after repeated heating and cooling.
These solutions exhibit the
periodicity in $~\mu^{\prime}~$ as discussed before.
For $~\mu^{\prime} \geq 1.5$ we have seen also
other branches of solutions with higher action than the
lowest possible one which are metastable under cooling.

In addition to the full action, the magnetic
(open squares) and electric contributions (full dots) to the action
are shown separately. The most interesting for us are the states
reached at half-integer and integer $~\mu'~$, respectively.
For $ \mu' \simeq 0.5, 1.5, \cdots $ we reproduce a BPS dyon state
identical to the original one at $ \mu' = 0.5 $ with equal
electric and magnetic contributions to the action.
For $ \mu' \simeq 1.0, 2.0, \cdots $ only the magnetic part of
the action survives. That means that we have obtained purely magnetic,
HP-like monopoles.

We convinced ourselves that we have obtained static solutions with
localized action and topological charge (the latter for the BPS dyon only).
As an example we show the action density profile for the case
$~\mu' = 1.0~$ being a pure HP-monopole (see Fig. 2).

In order to see what seems to be important of the boundary fields for the
classes of lattice solutions we have found and for the classification below
of the quantum fields on these background fields we compute the Polyakov 
line.  Fig. 3 shows the Polyakov line averaged over
spatial points $\vec{x}$ on 
the boundary, where the fields are kept at classical (discretized)
BPS dyon values, as a function of $\mu'$. We see the periodicity
in $~\mu'~$ again. For integer values $\mu'$, where inside the box
HP monopoles are supported as classical solutions, we have $~<L> \simeq
\pm 1~$. On the contrary for half-integer $~\mu'~$ providing BPS dyon
solutions $~<L> \simeq 0~$. (Small deviations of the positions of the minima
from $0$ and $\pm 1$ are due to an uncomplete
optimization of the $R_{eff}$-value on $N_s$ in the definition of
$\mu'$.)

\section{Quantum Fields with Dyon Boundary \\ Conditions}
In the following we want to investigate quantum fields with fixed spatial
boundary conditions as given by the ansatz (\ref{dyon1}, \ref{dyon2}).
We call the latter simply {\it dyon boundary conditions}. But we
should keep in mind that these boundary conditions for appropriate
values of $~\mu'~$ are compatible to classical BPS dyon or HP monopole
solutions of the lattice equations of motion. We emphasize that these
solutions are stable. This provides the opportunity to study a real
semi-classical situation with a unique background field parametrized
by $~\mu'$.

All our present work refers to $SU(2)$ pure gauge theory on a thermal
(non-symmetric) lattice as a function of $\beta$
and $~\mu'~$. The simulations have been done with a standard Monte Carlo
Metropolis algorithm.
For comparison, we have also done simulations at the same $\beta$ values
on the same lattices with periodic boundary conditions.
For the simulations, the lattice size has been $~16^3 \times 4~$.
Remember that for $~N_t = 4~$ the deconfinement phase transition occurs at
$~\beta = \beta_c \simeq 2.29~$ \cite{engels}.
In the following we will mainly consider two typical cases:
$\beta=2.2$, characteristic for the confinement phase, and $\beta=2.4$,
for the deconfinement phase.

In order to estimate the range of influence of the dyon boundary condition
on the quantum fields we computed different local observables for various
$~\mu'$ as a function of the (minimal) distance $~d~$ from the lattice site
to the boundary. Figs. 4 and 5 show the plaquette
contributions to the magnetic and electric part of the action averaged over
all plaquettes with a given distance $~d~$. The situation for both
typical $\beta$-values looks quite similar. Inside the box we obtain
plateau values. The range of the plateau seems to define a core size
of 'normal' quantum fluctuations. We shall investigate the properties
of the quantum fields inside the core in detail. There at least
no difference is seen between magnetic and electric contributions to the
action.

We have computed the average Polyakov
line as a function of $~d~$, too. This quantity is able to tell more
about the influence of the type of boundary condition on the quantum fields
in the interior of the box. The results are plotted in
Figs. 6 and 7, respectively. Even inside the core
established before (roughly at $~d \geq 3$) the behaviour of the
Polyakov line strongly depends on the boundary conditions compatible
either with BPS dyon or with HP monopole background fields. For integer
$\mu'$ (HP monopole) in both the confinement and the deconfinement phase
the results are compatible with the one obtained for periodic boundary
conditions at the same $\beta$. That means that the interior of the
lattice is in the phase corresponding to the $\beta$ value.
For $\beta = 2.2$, in the presence of monopole boundary conditions, the
small deviation (in the center of the lattice) from the
expected zero is only due to the maximal violation of the $Z(2)$ invariance
right on the spatial boundary (compare with Fig. 3).

For $\beta = 2.4$ (Fig. 7) we show the average Polyakov line
for periodic boundary conditions in a
symmetrized way. In fact, during our relatively short simulation runs of
typically $O(1000)$ configurations with $50$ empty sweeps
we did not observe any tunneling between the $Z(2)$-symmetric states.
In the presence of monopole boundary conditions, going deeper into the
lattice the locally averaged Polyakov line approaches the two
$Z(2)$-symmetric values.
On the contrary, for half-integer $\mu'$ (BPS dyon) the average
Polyakov line remains zero throughout the whole lattice
when passing the deconfinement transition.
At $\beta = 2.4$ inside the core we clearly get $|<L>| < 0.1$.
These observations are supported by the corresponding histograms of local
Polyakov line values measured at distance $~d = 5~$ from the boundary
which are shown in Figs. 8 and 9 for
$\beta = 2.2$ and $2.4$, respectively.

In the confinement phase, the histograms corresponding to the monopole
b.c. are displaced representing  entirely the effect of the finite
correlation length of the Polyakov line, while the histogram for the dyonic 
boundary conditions coincides with the 'normal' one (for periodic boundary
conditions).
In the deconfinement phase the normal histogram (without tunneling
as in our case) is reproduced with monopole boundary conditions at
distance $~d = 5~$, while the histogram for dyonic boundary conditions
is almost symmetric, similar to the lower $\beta$-value.
We conclude that, while the HP monopole b.c. are
compatible also with the confinement phase, a BPS dyon background
inside the finite box keeps the system
in the confinement state even for temperatures above $T_c$.
But as we have seen for smaller $N_t$ or larger $\beta$,
this does not persist at arbitrarily high temperature.
We checked this for $~N_t = 2,~~\beta = 2.4~$
(compare Fig. 10) and $~N_t = 4,~~\beta = 2.7~$.
In both cases of boundary conditions we have obtained Polyakov
averages inside the corresponding core regions, being compatible
with the case of periodic boundaries, i.e. deconfinement states.
Therefore, at larger temperatures deconfinement is restored inside the
box in spite
of the dyon boundary condition.

\section{Conclusions}
We have investigated pure $SU(2)$ lattice gauge theory at finite
temperatures. We have imposed fixed spatial boundary conditions defined
by a discretized static Bogomol'nyi-Prasad-Sommerfield monopole or dyon
solution of the continuum Yang-Mills field equations. By minimizing the
lattice action we have found stable field configurations which solve the
lattice equations of motion. As already reported by Smit and van der Sijs
\cite{smitvds1}
these configurations can be self-dual BPS-like dyon  or pure magnetic
HP-like monopole solutions. They occur periodically in terms
of the core size of the original dyon as fixed at the spatial boundary.

As far as these configurations are really stable, i.e. repeated cycles of
Monte Carlo heating and subsequent cooling provide always the same result,
we are in a position to simulate the quantum theory in well-defined
classical background fields.

We have used this environment to find out, how dyon-like or
purely magnetic, monopole-like
background fields influence the finite temperature quantum fields.
Our main observation was that in the dyon-case the fields inside
the box are kept in the confinement state even for higher temperatures,
which in the standard case of periodic boundary conditions cause
deconfinement. The Polyakov line as the main order parameter for the
deconfinement transition turned out to be zero remarkably stable
inside the whole $3D$ volume considered.  However, a further
increase of the temperature finally restored the deconfinement inside the
$3D$  box. The phenomenon reported here might be called {\it delay
of confinement evaporation} due to dyon boundary conditions.

We would like to interprete our findings such that the confinement phenomenon
is stronger related to self-dual semi-classical objects like BPS dyons
than to a pure magnetic HP monopole background.
On the other hand, deconfinement, when described semi-classically, requires
background fields which are not (anti-) self-dual.
The final loss of confinement above $\beta=2.7$ means breakdown of
semi-classical approximation.
This supports the view
developed in recent papers, where the correlation between instantons
and monopoles has been investigated in detail on the lattice
(see \cite{inst_mon}). The idea that dyons might
provide confinement within a semi-classical framework has been
put forward also within the continuum approach \cite{superconf}.

Our results were obtained for quite small lattice sizes. We are going to
check them deeper in the continuum limit and for physically larger box
volumes.

\section*{Acknowledgements}
The authors are grateful to B.V. Martemyanov, A. van der Sijs,
Yu.A. Simonov, J. Smit and M.I. Polikarpov for useful discussions.
This work was partly supported by RFBR grants N 97-02-17491
and N 96-02-17230, INTAS-RFBR 95-0681  and INTAS 96-370.
The authors acknowledge also the support by the joint DFG-RFFI
grant 436 RUS 113/309/0 (R) or RFBR-DFG grant N 96-02-00088 G.
M. M.-P. acknowledges partial support by the EC TMR network 
FMRX-CT97-0122.

%
\pagestyle{empty}
%
%
\begin{figure}[htb]
\vspace*{-1.0cm}
\begin{center}
\hbox{
\epsfysize=16cm\epsfxsize=16cm
\epsfbox{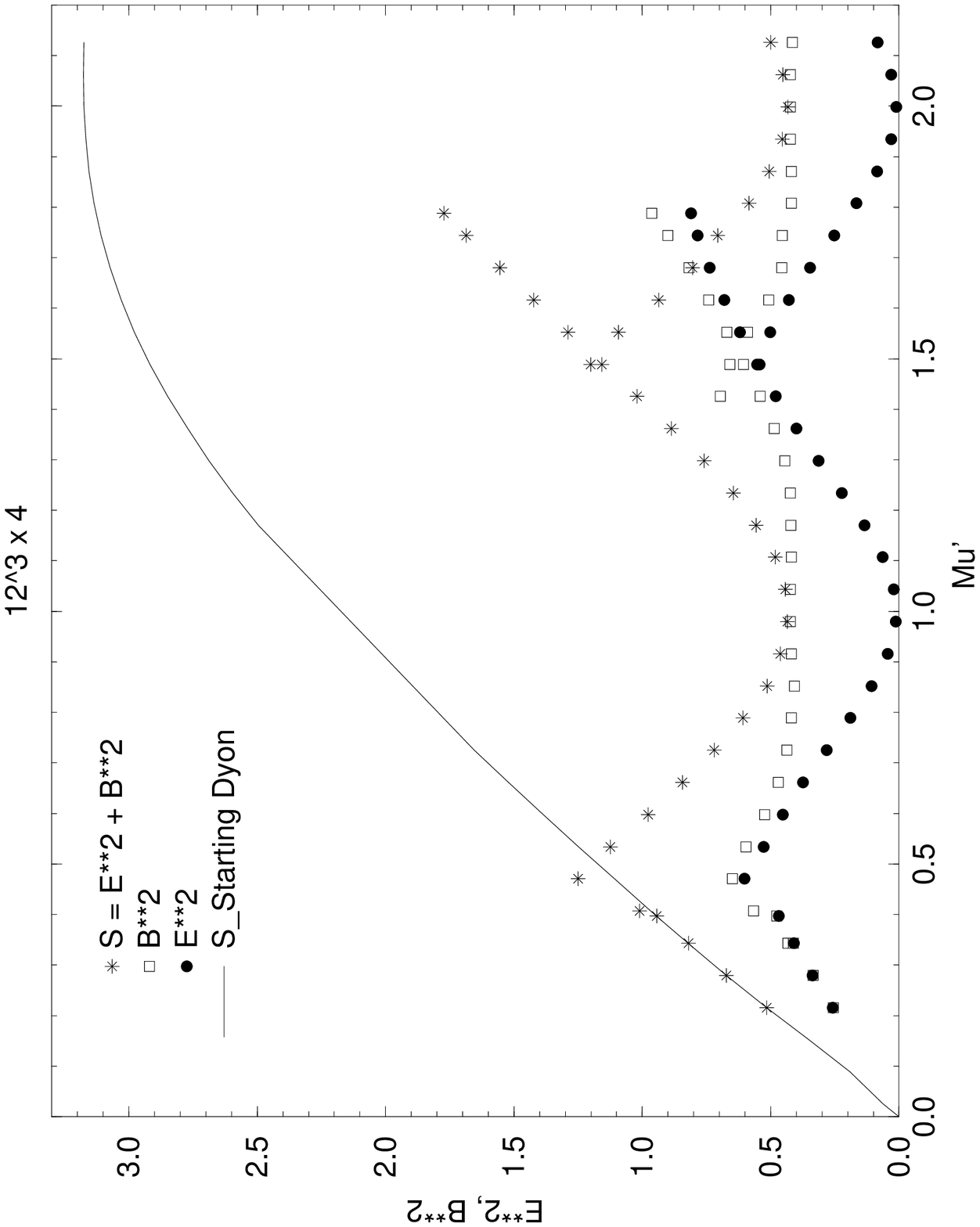}
     }
\vspace*{1.0cm}
\caption{
Total action, magnetic and electric part of the action
for possible classical solutions of the lattice field equations with
dyon boundary conditions as explained in the text.
 }
\end{center}
\label{fig1}
\end{figure}
%
%
\begin{figure}[htb]
\vspace*{-1.0cm}
\begin{center}
\hbox{
\epsfysize=16cm\epsfxsize=16cm
\epsfbox{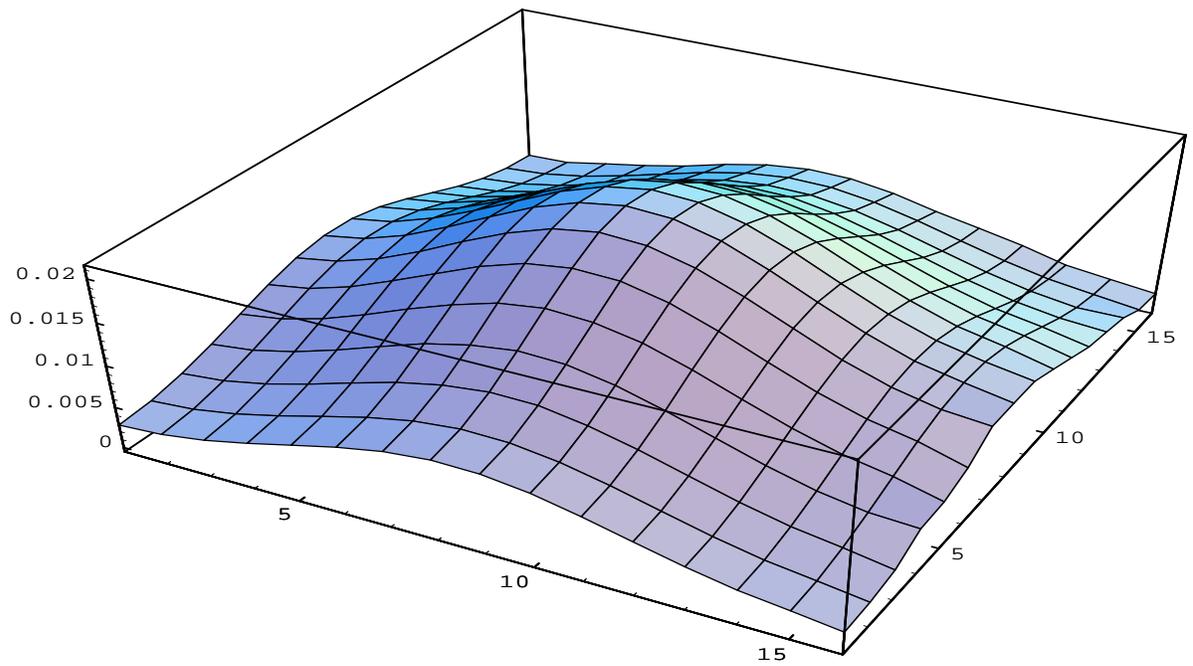}
     }
\vspace*{1.0cm}
\caption{
Action density profile of the classical solution of the lattice equations
of motion for $\mu'=1.0$ (pure magnetic monopole) as a function of
$~y = 1,\cdots,16~$ and $~z = 1,\cdots,16~$.
 }
\end{center}
\label{fig2}
\end{figure}
%
%
\begin{figure}[htb]
\vspace*{-1.0cm}
\begin{center}
\hbox{
\epsfysize=16cm\epsfxsize=16cm
\epsfbox{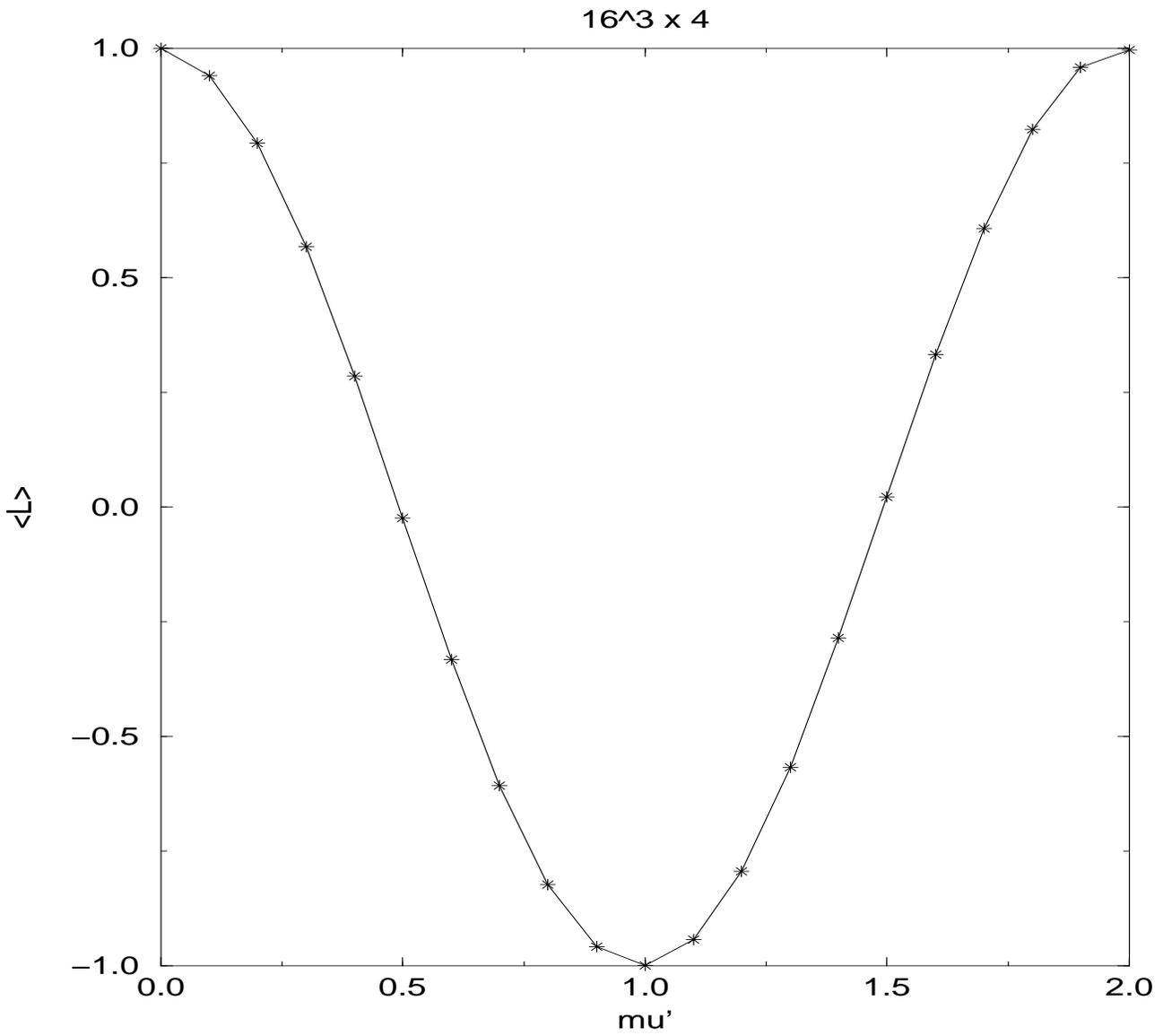}
     }
\vspace*{1.0cm}
\caption{
Polyakov line operator computed for the classical BPS configuration
at the boundary of the 3D box for varying parameter $\mu'$.
 }
\end{center}
\label{fig3}
\end{figure}
%
%
\begin{figure}[htb]
\vspace*{-1.0cm}
\begin{center}
\hbox{
\epsfysize=16cm\epsfxsize=16cm
\epsfbox{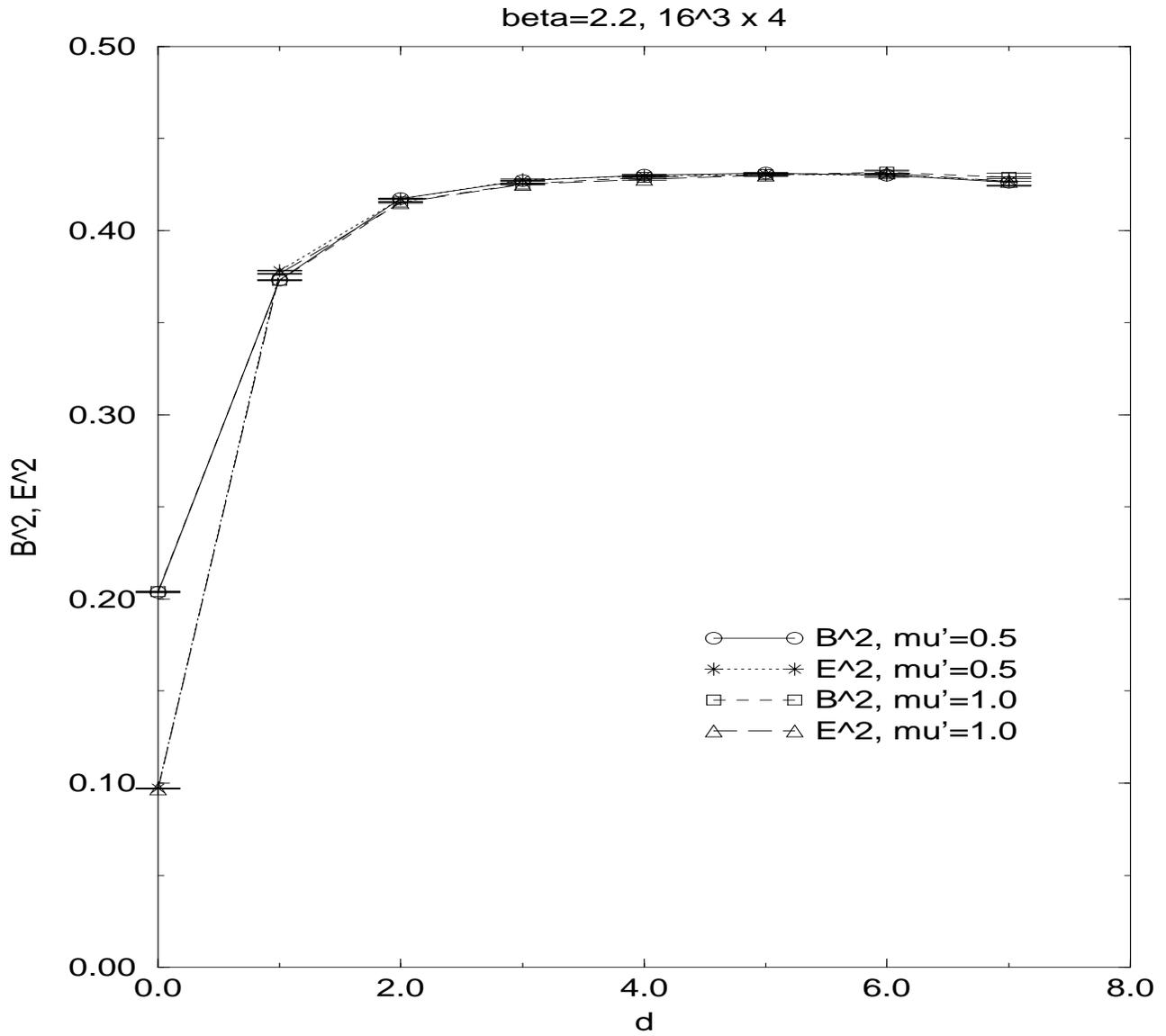}
     }
\vspace*{1.0cm}
\caption{
Magnetic and electric contributions to the full action from pla\-quettes with
distance $~d~$ from the boundary in the dyon ($\mu'=0.5$)
and in the pure monopole case ($\mu'=1.0$) for $\beta = 2.2$ (confinement).
 }
\end{center}
\label{fig4}
\end{figure}
%
%
\begin{figure}[htb]
\vspace*{-1.0cm}
\begin{center}
\hbox{
\epsfysize=16cm\epsfxsize=16cm
\epsfbox{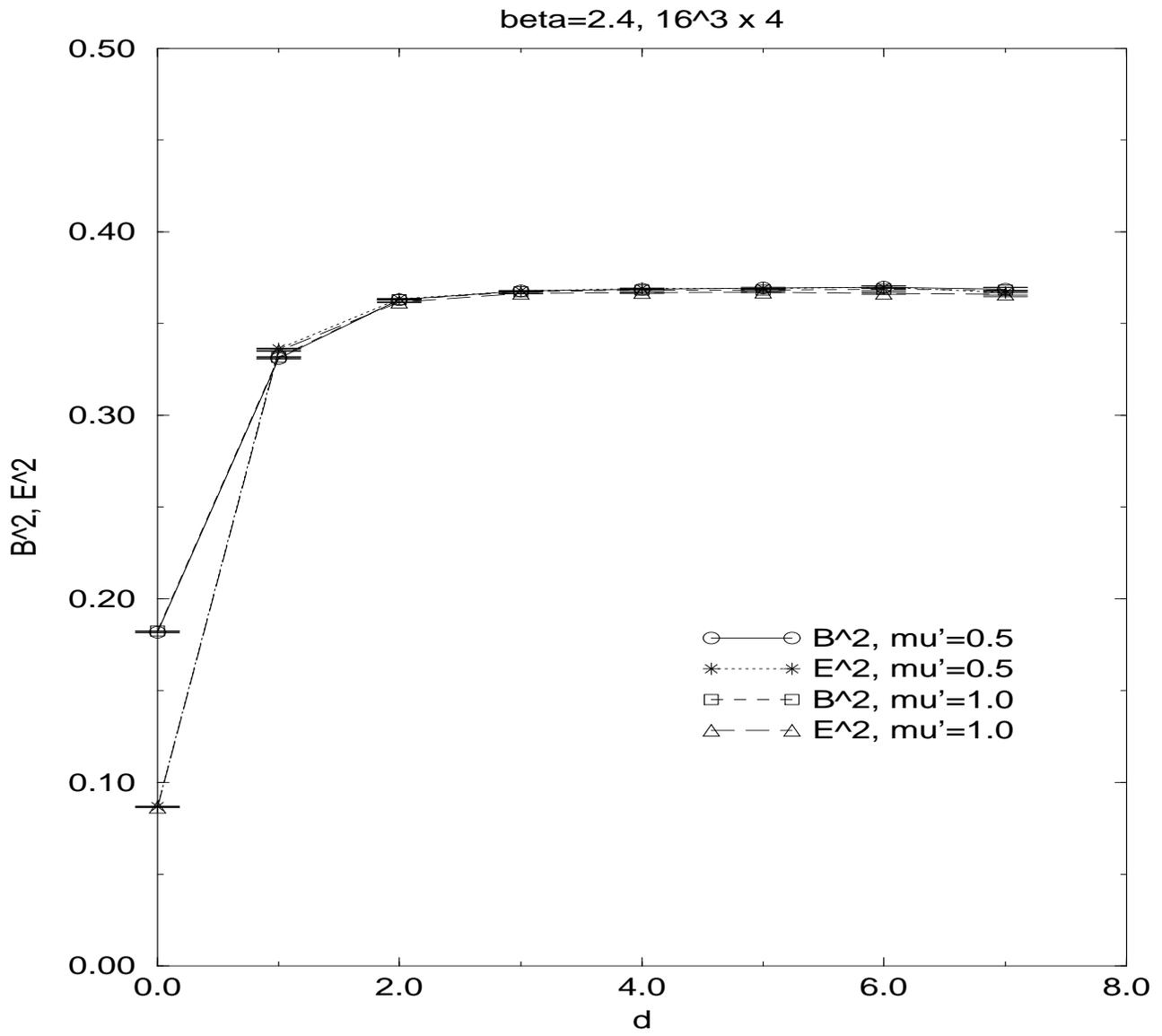}
     }
\vspace*{1.0cm}
\caption{
As in Fig. 4, but for $\beta = 2.4$ (deconfinement).
 }
\end{center}
\label{fig5}
\end{figure}
%
%
\begin{figure}[htb]
\vspace*{-1.0cm}
\begin{center}
\hbox{
\epsfysize=16cm\epsfxsize=16cm
\epsfbox{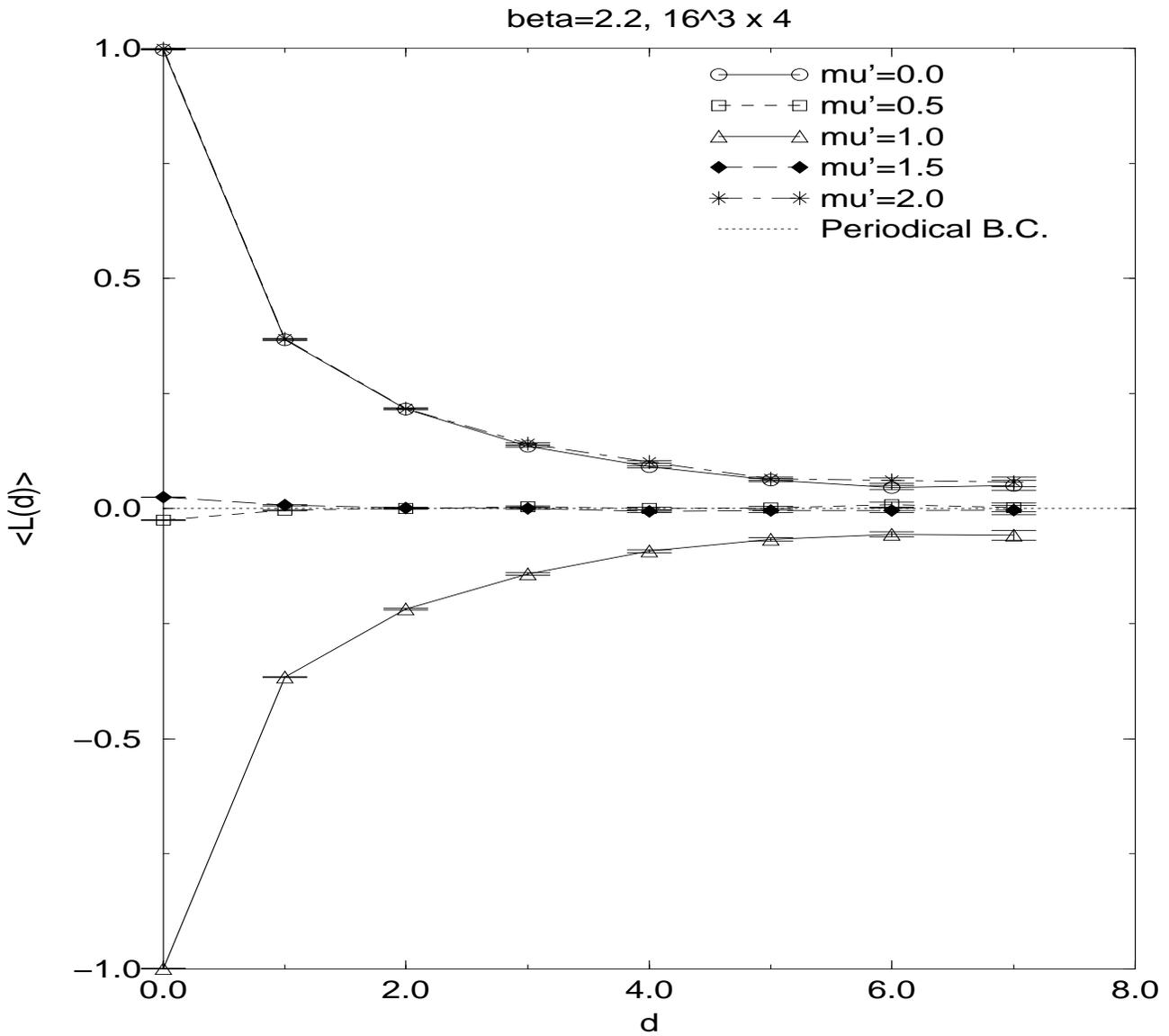}
     }
\vspace*{1.0cm}
\caption{
Average Polyakov line operator for $\beta=2.2$ (confinement phase)
computed for different boundary conditions
at all lattice sites with distance to the boundary $~d~$.
For comparison we show the result obtained with periodic
boundary conditions (dotted lines), too.
 }
\end{center}
\label{fig6}
\end{figure}
%
%
\begin{figure}[htb]
\vspace*{-1.0cm}
\begin{center}
\hbox{
\epsfysize=16cm\epsfxsize=16cm
\epsfbox{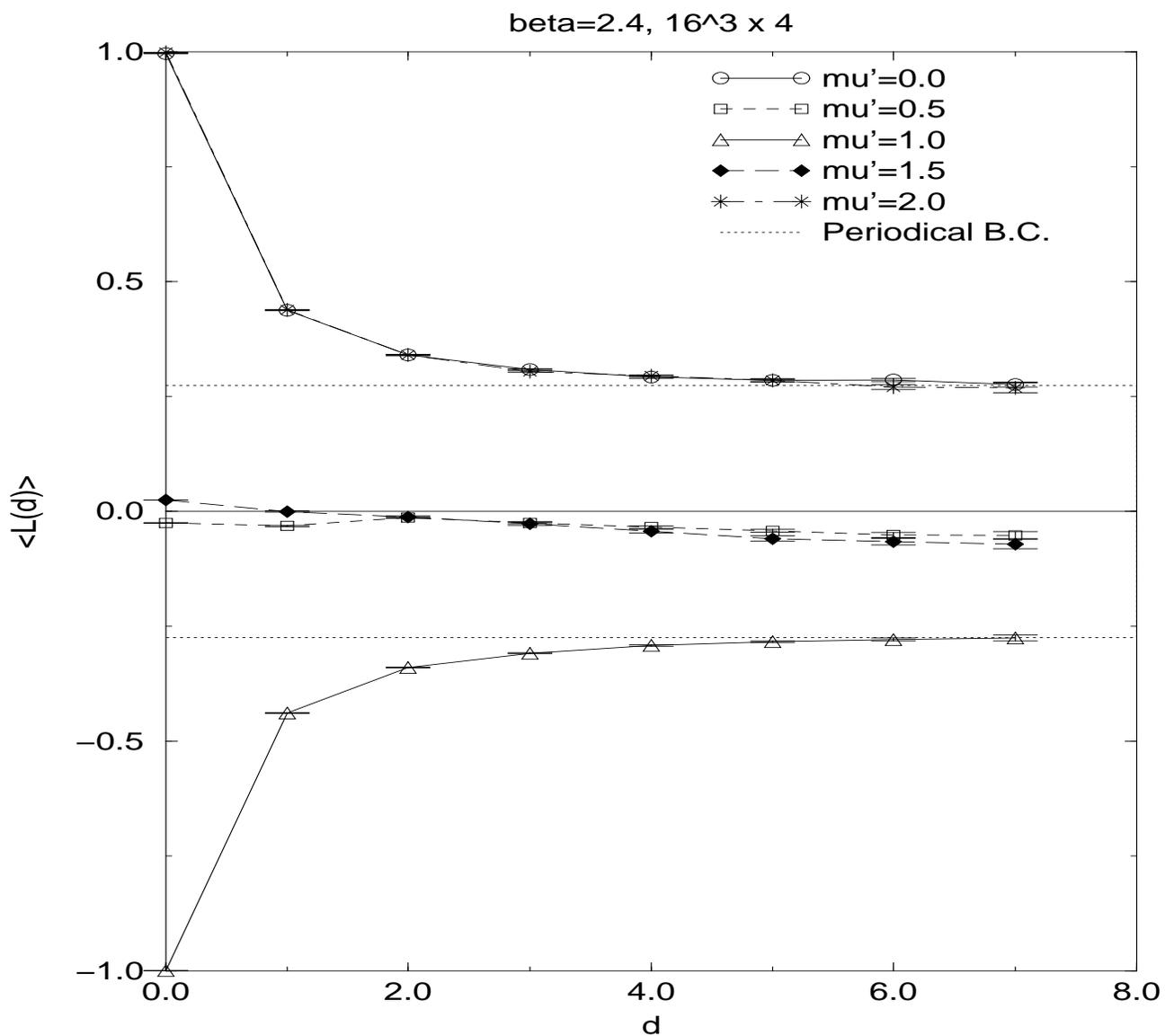}
     }
\vspace*{1.0cm}
\caption{
As in Fig. 6, but for $\beta=2.4$ (deconfinement phase).
The results obtained with periodic
boundary conditions were splitted into the two $Z(2)$ asymmetric values
(dotted lines).
 }
\end{center}
\label{fig7}
\end{figure}
%
%
\begin{figure}[htb]
\vspace*{-1.0cm}
\begin{center}
\hbox{
\epsfysize=16cm\epsfxsize=16cm
\epsfbox{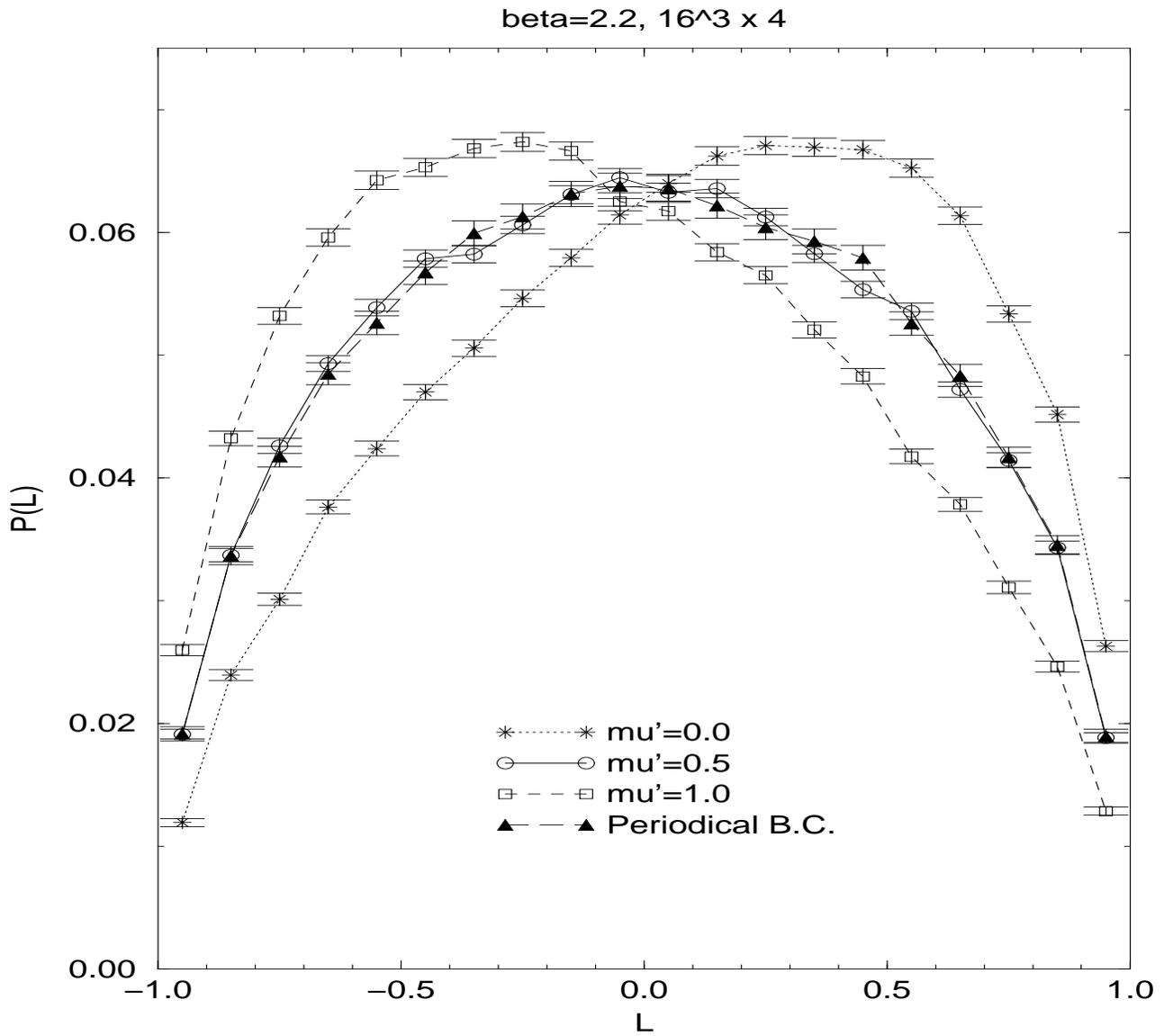}
     }
\vspace*{1.0cm}
\caption{
Distribution of local Polyakov line values for various boundary conditions
(dyon case for $\mu'=0.5$, pure magnetic monopole case $\mu'=0.0, 1.0$).
The Polyakov line operator has been computed at all sites with
distance $~d = 5~$ from the boundary. For comparison we show also the
case of periodic boundary conditions obtained at the same $~\beta=2.2~$.
 }
\end{center}
\label{fig8}
\end{figure}
%
%
\begin{figure}[htb]
\vspace*{-1.0cm}
\begin{center}
\hbox{
\epsfysize=16cm\epsfxsize=16cm
\epsfbox{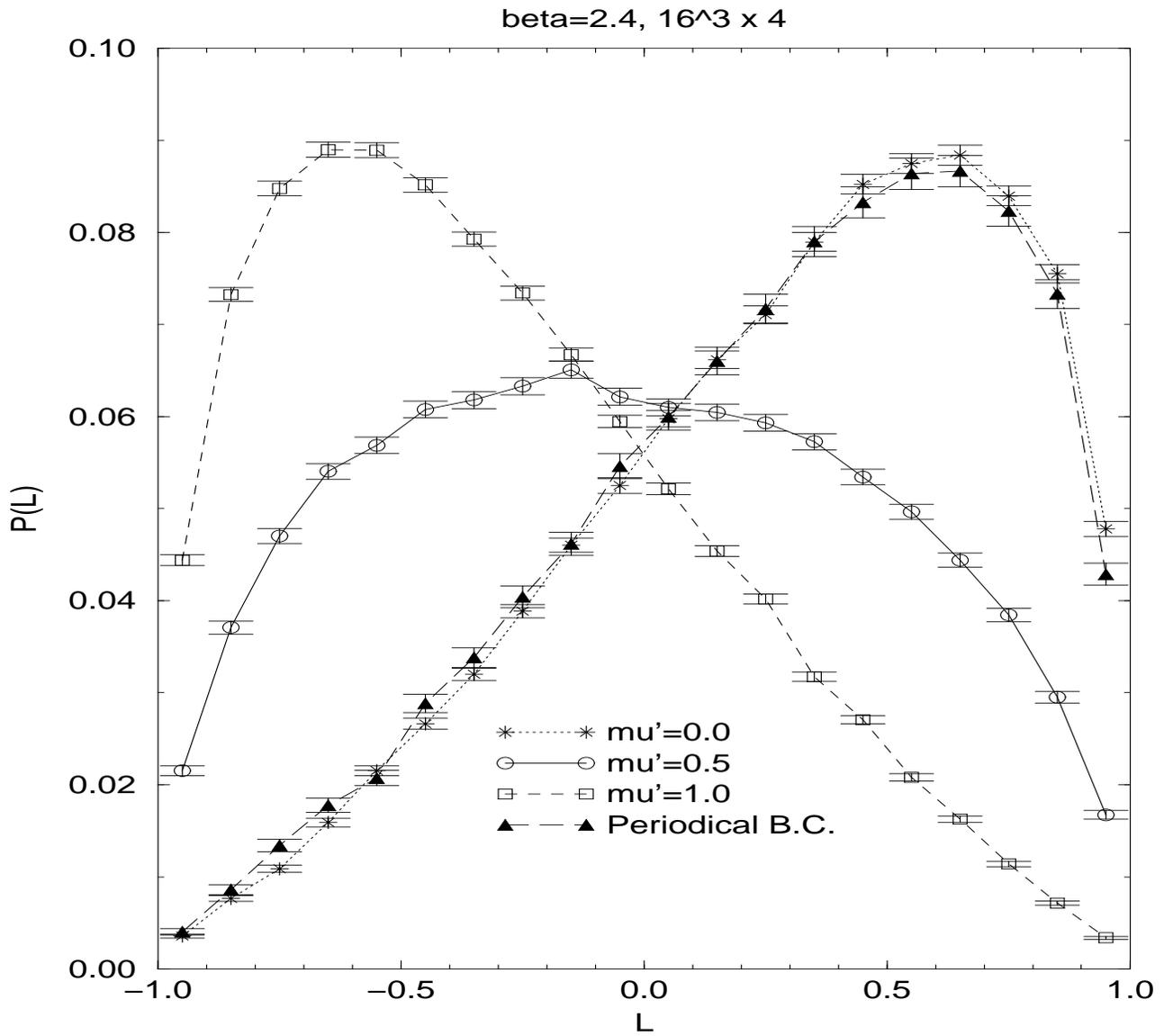}
     }
\vspace*{1.0cm}
\caption{
The same as in Fig. 8, but for $~\beta = 2.4~$.
All runs were taken with
comparable statistics ($O(500)$ configurations). In the periodic boundary
case, this statistics was not enough in order to observe any tunneling
between the $Z(2)$ states. Therefore, the distribution occurs to be
asymmetric confined to only one of the $Z(2)$ orientations.
 }
\end{center}
\label{fig9}
\end{figure}
%
%
\begin{figure}[htb]
\vspace*{-1.0cm}
\begin{center}
\hbox{
\epsfysize=16cm\epsfxsize=16cm
\epsfbox{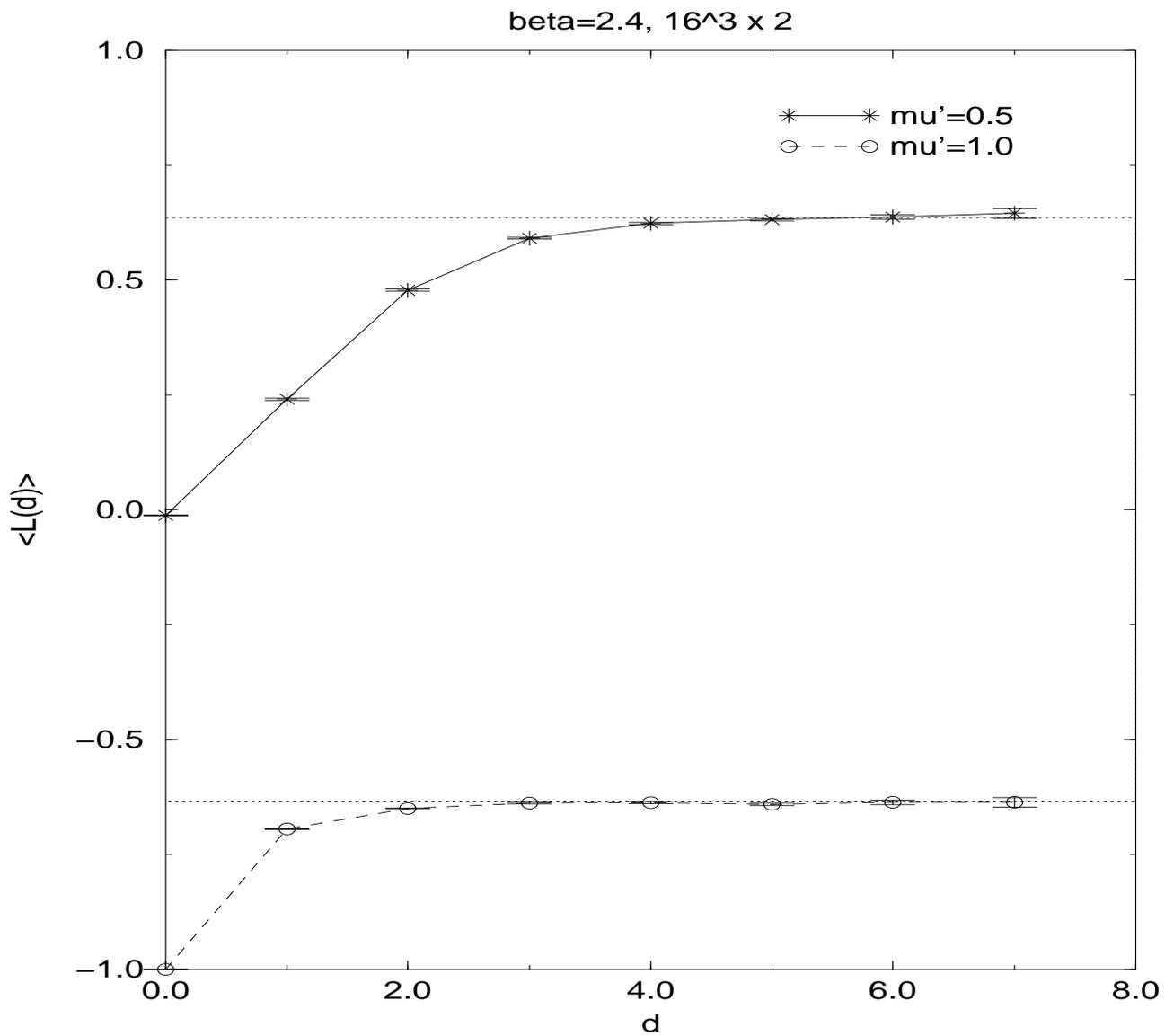}
     }
\vspace*{1.0cm}
\caption{
As for Fig. 7, but for higher temperature $~N_t = 2~$.
 }
\end{center}
\label{fig10}
\end{figure}
\end{document}